\documentclass[aps,pra,twocolumn,showpacs,amsmath,amssymb,floatfix,superscriptaddress]{revtex4}
\usepackage{epsfig}

\usepackage{bm}
\usepackage{color}
\usepackage{ulem}

\begin{document}

\title{Kerr non-linearity in a superconducting Josephson metamaterial}

\author{Yu. Krupko}
\affiliation{Univ. Grenoble Alpes, CNRS, Grenoble INP, Institut N\'eel, 25 rue des Martyrs BP 166, 38042 Grenoble, France}
\author{V. D. Nguyen}
\affiliation{Laboratoire de Physique et Mod\'elisation des Milieux Condens\'es,
Universit\'e Grenoble Alpes and CNRS,
25 rue des Martyrs, 38042 Grenoble, France}
\author{T. Wei\ss l}
\affiliation{Univ. Grenoble Alpes, CNRS, Grenoble INP, Institut N\'eel, 25 rue des Martyrs BP 166, 38042 Grenoble, France}
\affiliation{Nanostructure Physics, Royal Institute of Technology (KTH), Roslagstullbacken 21, SE-10691 Stockholm, Sweden}
\author{\'E. Dumur}
\affiliation{Institute for Molecular Engineering and Materials Science Division, Argonne National Laboratory, Argonne, IL}
\affiliation{Univ. Grenoble Alpes, CNRS, Grenoble INP, Institut N\'eel, 25 rue des Martyrs BP 166, 38042 Grenoble, France}
\author{J. Puertas}
\affiliation{Univ. Grenoble Alpes, CNRS, Grenoble INP, Institut N\'eel, 25 rue des Martyrs BP 166, 38042 Grenoble, France}
\author{C. Naud}
\affiliation{Univ. Grenoble Alpes, CNRS, Grenoble INP, Institut N\'eel, 25 rue des Martyrs BP 166, 38042 Grenoble, France}

\author{F.~W.~J.~Hekking}
\thanks{$\dagger$ Deceased 15th Mai 2017}
\affiliation{Laboratoire de Physique et Mod\'elisation des Milieux Condens\'es,
Universit\'e Grenoble Alpes and CNRS,
25 rue des Martyrs, 38042 Grenoble, France}

\author{D. M. Basko}
\affiliation{Laboratoire de Physique et Mod\'elisation des Milieux Condens\'es,
Universit\'e Grenoble Alpes and CNRS,
25 rue des Martyrs, 38042 Grenoble, France}

\author{O. Buisson}
\affiliation{Univ. Grenoble Alpes, CNRS, Grenoble INP, Institut N\'eel, 25 rue des Martyrs BP 166, 38042 Grenoble, France}
\author{N. Roch}
\affiliation{Univ. Grenoble Alpes, CNRS, Grenoble INP, Institut N\'eel, 25 rue des Martyrs BP 166, 38042 Grenoble, France}
\author{W. Hasch-Guichard}
\affiliation{Univ. Grenoble Alpes, CNRS, Grenoble INP, Institut N\'eel, 25 rue des Martyrs BP 166, 38042 Grenoble, France}

\date{\today}

\begin{abstract}
We present a detailed experimental and theoretical analysis of the dispersion and non-linear Kerr frequency shifts of plasma modes in a one-dimensional Josephson junction chain containing 500 SQUIDs in the regime of weak nonlinearity. The measured low-power dispersion curve agrees perfectly with the theoretical model if we take into account the Kerr renormalisation of the bare frequencies and the long-range nature of the island charge screening by a remote ground plane. We measured the self- and cross-Kerr shifts for the frequencies of the eight lowest modes in the chain. We compare the measured Kerr coefficients with theory and find good agreement.
\end{abstract}

\maketitle

\section{Introduction}
Metamaterials have artificially engineered properties that do not occur in nature and enable one to control interactions of matter with electromagnetic waves. Here a particular interest relies in the realisation of negative or high refractive index materials~\cite{Plourde,Eleftheriades, Anlage, Pendry, Jung,Alu}. Superconducting circuits operating in the microwave region such as Josephson junction chains offer a unique possibility for the design of metamaterials since electromagnetic signals can propagate in such circuits with extremely low losses, the circuit properties can
be tuned by applying an external magnetic field, and the Josephson effect provides a mechanism for strong non-linearity. Numerous applications of superconducting metamaterials range from amplifiers and detectors to quantum information and metrology~\cite{Jung, Guichard}.

Here we present Kerr effect measurements in a Josephson junction chain containing 500 SQUIDs (Superconducting Quantum Interfence Devices), and find a non-linear relative change of the refractive index $\Delta n/n\sim {10}^{-7}$ per photon, 11 orders of magnitude larger than typically observed in optical systems~\cite{Boulanger}. We perform a detailed comparison between our measurements and theory. We also characterise the dispersion relation of the linear waves in the chain, and demonstrate the necessity to include the long-range nature of the island charge screening by a remote ground plane for its quantitative understanding.

Our results apply directly to the realisation of parametric amplifiers at the quantum limit of noise based on Josephson junction chains \cite{Macklin, White, Planat,Castellanos,Yurke}. We expect as well a potential use of our superconducting metamaterial in the realisation of quantum simulations based on superconducting circuits \cite{Houck, Le Hur}.

Josephson junction chains have been studied for more than three decades motivated intially as a model system for the study of the zero-temperature superconductor-to-insulator quantum phase transition in superconducting granular films~\cite{Efetov, Bradley}. The superconductor-to-insulator quantum phase transition has been observed in granular films~\cite{Jaeger} and wires~\cite{Giordano}, as well in Josephson junction chains~\cite{Chow, Haviland}. More recently the superconductor-to-insulator quantum phase transition of long Josephson junction chains has regained interest and the nature of the insulating state was studied~\cite{Duty}. Quantum phase-slips have been studied in Josephson junction chains \cite{Rastelli, Erguel, Pop2}. Other recent experiments successfully employed Josephson junction chains as a high inductance environement for quantum systems such as superconducting qubits~\cite{Pop,Masluk} or quantum conductors~\cite{Altamiras}. These chains were also suggested theoretically as a platform for the study of the dynamics of the spin-boson model realised in a superconducting circuit which couples a superconducting qubit to a high linear impedance environment\cite{Bera, Le Hur, Goldstein, Puertas}. Nonlinear effects occuring in Josephson junction chains might be used as well for the generation of non-classical states of microwaves\cite{Fistul,Imamoglu,Bourassa}.

In this article we present a detailed experimental and theoretical analysis of the dispersion and non-linear Kerr frequency shifts of plasma modes in a one-dimensional Josephson junction chain containing 500 SQUIDs in the regime of weak nonlinearity. 
The article is structured as follows. In Section II we give a description of the sample and of the experiment. Section III summarizes the theory of a weakly non-linear chain taking into account for long range Coulomb interactions in the chain. 
In Section IV we present the measurements of the dispersion relation of a 500 SQUID chain and compare our measurements to our theory. Section V shows our results on the measured self- and cross-Kerr coefficients in the regime of weak non-linearity and we compare them to the theoretical expectations and find good agreement. Finally, Section VI presents a conclusion and an outlook of our results.

\section{Sample and experiment description}

\begin{figure}[]
\includegraphics[width=8.5cm]{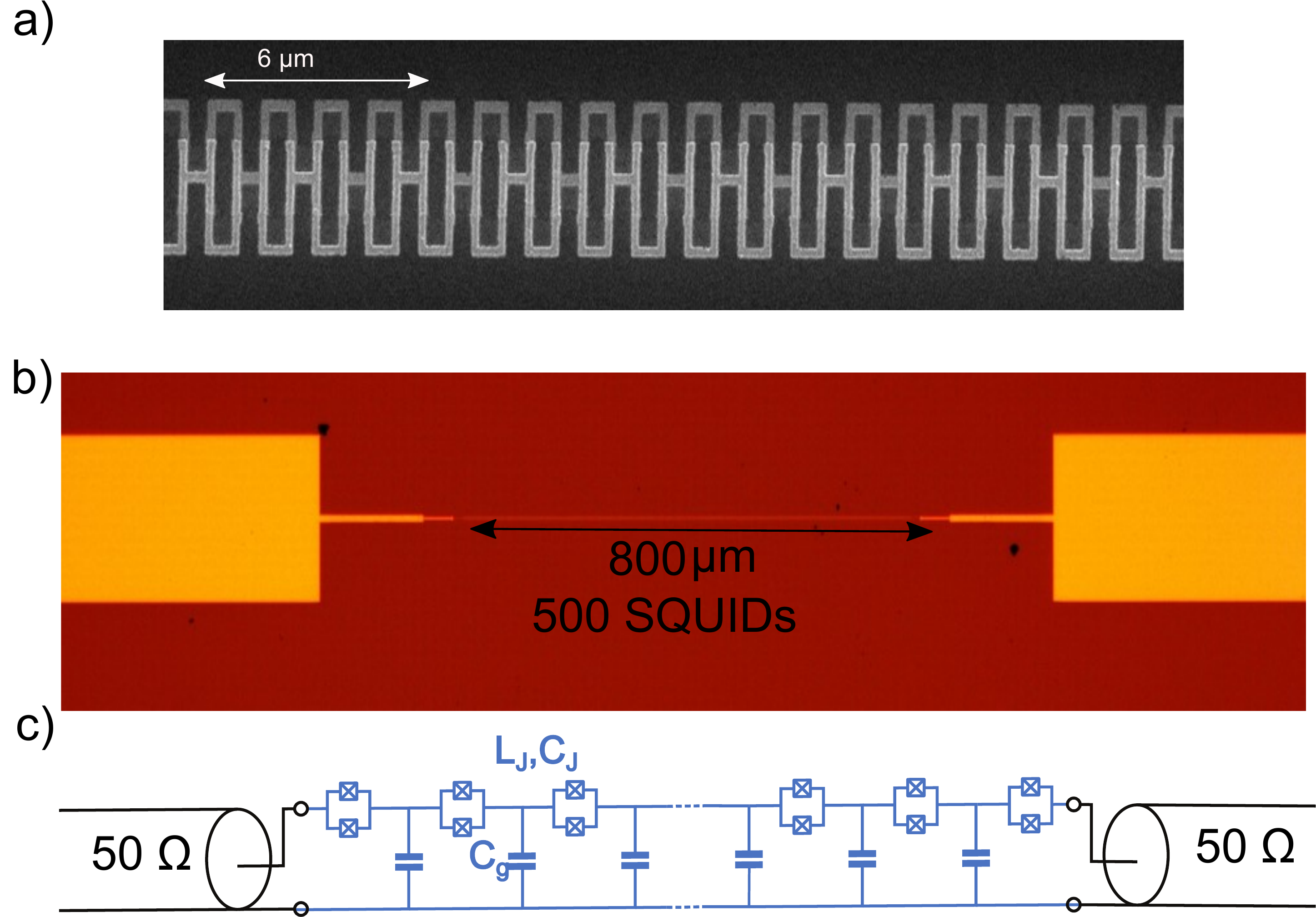}
\caption{(a) SEM image of one part of the 500-SQUID chain. (b) Optical image showing the SQUID chain connected at both ends to a microstrip transmission line. (c) Equivalent electrical scheme of the chain and transmission line for the local screening model.}
\label{sample}
\end{figure}

We fabricated one-dimensional chains of SQUIDs by shadow evaporation of aluminum on a 300$\mu$m thick high-resistivity silicon substrate. A 100~keV electron beam lithography system and a bridge-free technic based on an asymmetric undercut \cite{Lecocq} were employed to prepare the resist masks. Figs.\ \ref{sample} (a) and (b) show SEM and optical images of our chain containing 500 SQUIDs. The chain is connected at both ends to a 50$\Omega$ microstrip transmission line, which was fabricated during the same fabrication step as the junctions (see Appendix~\ref{App:ExpTech} for the details of the measurement technique). The Josephson junction area associated with each SQUID is $0.72\:\mu\mbox{m}^2$. 
Using the value of $45\:\mbox{fF}/\mu\mbox{m}^2$ for junctions based on aluminium oxide \cite{Fay} we can deduce the capacitance for each SQUID as $C_J = 32.4$~fF. The corresponding charging energy is $E_C = e^2/2C_J$ leading to $E_C/h=0.598\:\mbox{GHz}$ where $h$ is the Planck constant. From room temperature measurements of the tunnel junction normal-state resistance, the Josephson energy can be estimated to be $E_J/h \simeq 104$~GHz, resulting in a ratio between the Josephson energy and the charging energy of $E_J/E_C \simeq 170 $.

Fig.\ \ref{sample}(c) represents the standard equivalent electrical scheme~\cite{FazioReview} of the Josephson junction chain connected at both ends to a $50\:\Omega$ transmission line. Here $C_J$ is the SQUID capacitance, $L_J=\hbar/(2eI_c\cos\phi)$ is the non-linear Josephson inductance of a SQUID, where $I_c$ is the critical current and $\phi$ the superconducting phase difference over the SQUID. $C_g$ is the capacitance of the superconducting island between two SQUIDs to the ground-plane. The ground plane is defined by the evaporation of a 200~nm thick gold layer on the back of the silicon substrate.

Two kinds of samples have been measured. Identical chains, consisting of 500 Josephson junctions, were coupled differently to the transmission line: one was embedded into the microwave strip line, as explained above (Fig.\ \ref{sample}), while the other one was coupled capacitively to the transmission line. The latter configuration, discussed in Appendix~\ref{App:A}, enables us to infer the internal quality factor of the chain modes. As it is easier to model theoretically the direct coupling of the Josephson junction chain to the microstrip line in terms of the calculation of the Kerr coefficients this design was used to study quantitatively the Kerr effect occurring in the chain, which is presented in Section~V.

\section{Theory of a weakly-non-linear Josephson junction chain with a remote screening gate}

The theory of self- and cross-Kerr effects between modes propagating along Josephson junction chains in the regime of weak nonlinearity was reported recently in detail in Ref.~\cite{Weissl2}. The regime of strong non-linearity producing bistable behaviour in a Josephson junction chain has been studied in Ref.~\cite{Muppalla}. Also the Kerr effect of modes of a transmission line resonator embedding a single Josephson junction has been studied theoretically\cite{Bourassa}. In this system only specific modes of the resonator undergo a Kerr effect depending whether the single junction is placed on a node or an anti-node of the resonators eigen mode. In a Josephson junction chain, all chain  modes undergo a Kerr effect as the non-linearity is distributed over the whole chain.  Moreover widely tunable positive and negative Kerr coefficients have been studied in a chain of asymmetric superconducting quantum interference devices with nearest-neighbor coupling through common Josephson junction\cite{Zhang}. 

The theory developed in Ref.~\cite{Weissl2} focuses on Josephson junction chains with a nearby screening ground plane, see Fig.~\ref{sample}(c). In this case the Coulomb interaction taken into account to describe the modes is short range and includes interaction between neighbouring islands as well as between islands and the ground plane. Then the standard way to model Josephson junction chains is by introducing the junction capacitances $C_J$ and a ground capacitance $C_g$ for each island~\cite{FazioReview}. The experiments presented here are performed with a chain on a dielectric substrate with a significant thickness, such that it is separated from the screening ground plane by a distance that is of the order or much larger than the wavelength of the modes.  Electrostatically, this situation cannot be described with a single ground capacitance per island. In addition the experimentally measured dispersion relation does not compare well with the standard dispersion relation used in~\cite{Weissl2}. Therefore the long range Coulomb interaction between the islands of the chain must be included to describe the physics of the propagating modes. The following theoretical analysis goes beyond the usual models developped for Josephson junction chains (see Ref.~\cite{Weissl2}) and describes the dispersion relation in this situation of long range interaction of the modes as well as the Kerr and cross-Kerr coefficients. The coherence of these propagating modes has been studied in contexte with the superconductor-insulator quantum phase transition in Ref.~\cite{Kuzmin}.

\subsection{Hamiltonian}
The starting point of the theory is the well-known quantum phase model~\cite{FazioReview} for a Josephson junction chain with $N$ junctions connecting $N+1$ superconducting islands. Island $n$ carries a charge $Q_n$ and a phase $\phi_n$, with $n=0, \ldots, N$. Since in our sample both ends of the chain are connected to a low-impedance circuit, the appropriate boundary conditions are $V_0=V_N=0$ leading to $\phi_0=\phi_N=0$. Therefore, our problem reduces to $N-1$ degrees of freedom.  The chain is described by the Hamiltonian
\begin{equation}
\begin{aligned}
\hat{H} &= \frac{1}{2} \sum \limits_{n,m =1}^{N-1} \hat{Q}_n C^{-1}_{nm} \hat{Q}_m - E_J \sum \limits_{n=1}^{N-2} \cos(\hat{\phi}_{n+1} - \hat{\phi}_{n})\\
&-E_J\cos(\hat{\phi}_1)-E_J\cos(\hat{\phi}_{N-1}).
\end{aligned}
\label{eq:qpm}
\end{equation}
In this model, charge $\hat{Q}_n$ and phase $\hat{\phi}_m$ are conjugate variables, such that $[\hat{Q}_n,\hat{\phi}_m] = -2ie \delta_{n,m}$, with $e$ is the positive electron charge. 

The first term of the Hamiltonian is the charging energy. It describes the Coulomb interaction between charges on grains $n$ and $m$. It depends on the chain's inverse capacitance matrix $\widehat{C}^{-1}$ with matrix elements $C_{nm}^{-1}$. The form of this matrix is significantly different depending on whether the island charges are screened locally by the ground plane or not. The matrix is precisely determined by the electrostatic configuration of the chain with respect to nearby dielectrics and gates as well as by the boundary conditions. The second term is the total Josephson coupling energy of the chain. It is the sum of the nonlinear Josephson energies  $-E_J \cos(\hat{\phi}_{n+1}-\hat{\phi}_n)$ of neighboring islands $n$ and $n+1$, with a characteristic coupling strength $E_J$.

In the limit where the Josephson energy is much larger than the characteristic charging energy (we will provide a more detailed criterion below), the phase differences $\hat{\phi}_{n+1}-\hat{\phi}_n$ between neighboring islands are small, and the nonlinear Josephson energy can be expanded in powers of these phase differences. This is the weakly non-linear regime of interest here. Retaining the two lowest nonvanishing orders and dropping the constant term, we can approximate the Hamiltonian  as $\hat{H} \simeq \hat{H}_0 + \hat{H}_1$, where
\begin{equation}
\begin{aligned}
\hat{H}_0 &= \frac{1}{2} \sum \limits_{n,m =1}^{N-1} \hat{Q}_n C^{-1}_{nm} \hat{Q}_m + \frac{E_J}{2} \sum \limits_{n=1}^{N-2} (\hat{\phi}_{n+1} -\hat{\phi}_{n})^2\\&+\frac{E_J}{2}\hat{\phi}_{0}^2+\frac{E_J}{2}\hat{\phi}_{N-1}^2
\label{eq:h0}
\end{aligned}
\end{equation}
is the quadratic unperturbed Hamiltonian
and
\begin{equation}
\hat{H}_1 = -\frac{E_J}{24} \sum \limits_{j=1}^{N-2} (\hat{\phi}_{n+1} -\hat{\phi}_{n})^4
\label{eq:h1}
\end{equation}
is the nonlinear quartic correction term, that we will treat as a perturbation.
\\

\subsection{Charging energy: effect of a remote ground plane}
\label{CPRGP}

In the standard situation one assumes a close ground plane to provide an additional gate capacitance $C_g$ that screens the remaining charge on each island locally. Then, the capacitance matrix of the chain is given by
\begin{equation}
\begin{split}
 \widehat{C} &= \left(
\begin{array}{cccccc} 2C_J+C_g & -C_J & 0 & \hdots & &\\
-C_J  &  2C_J+C_g & -C_J &  0 & \hdots &\\
0 & -C_J &  2C_J+C_g & -C_J &  0  & \hdots \\
\vdots & 0 & \ddots &  \ddots & \ddots & \ddots \\
& & & & & \\
\end{array}\right).
\label{eq:capmat}
\end{split}
\end{equation}
This is an $(N-1) \times (N-1)$ tridiagonal matrix. The main diagonal contains elements $2C_J+C_g$. Only the first diagonals above and below the main one are non-zero and contain $-C_J$, reducing in this model the Coulomb interactions to the nearest neighbors. 

However, in the present experiment the ground plane is not close. Indeed, the chain is located on top of a dielectric (silicon) substrate, at a distance $d\simeq{300}\:\mu\mbox{m}$ away from the ground plane, while the space above the chain is filled with air/vacuum. Crucially, $d$ is of the order of or larger than the mode wavelength, which varies between 1.6~mm for the first mode down to 40~$\mu$m for the highest measured mode number 43. Therefore the screening by the ground plane of each charge ${Q}_m$ cannot be local. Instead, one has to properly account for the long-range part of the Coulomb potential. This is described in detail in Appendix~\ref{App:ChargEnergy}.

 We thus find the total capacitance matrix of the chain to be
\begin{widetext}
\begin{equation}
\begin{split}
 \widehat{C} &= \left(
\begin{array}{cccccc} 2C_J+C_{g,11} & -C_J+C_{g,12} & C_{g,13} & \hdots & &\\
-C_J+C_{g,21}  &  2C_J+C_{g,22} & -C_J+C_{g,23} &  C_{g,24} & \hdots &\\
C_{g,31} & -C_J+C_{g,32} &  2C_J+C_{g,33} & -C_J+C_{g,34} &  C_{g,35}  & \hdots \\
\vdots & \ddots & \ddots &  \ddots & \ddots & \ddots \\
& & & & & \\
\end{array}\right).
\label{eq:fullcapmat}
\end{split}
\end{equation}
\end{widetext}
Note that the full capacitance matrix is no longer tridiagonal, describing the long range Coulomb interactions along the chain. The matrix does not contain any zero elements, so the matrix inverse $C^{-1}_{nm}$ has to be obtained numerically. Below we will use the improved capacitance matrix of equation~(\ref{eq:fullcapmat}) to calculate the Kerr coefficients and to analyze the chain's experimentally measured dispersion relation. We emphasize that although the long-range screening model presented above looks significantly more complex than the standard model with local screening, it has only one unknown parameter~$a_0$ which is a short distance cut-off length of the Coulomb interaction (all the rest is known from the geometry). In the standard local model, the ground capacitance $C_g$ is usually treated as a fit parameter, so the number of fit parameters is effectively unchanged. We will see that including the long-range screening enables us to obtain good fits for the dispersion relation, whereas the use of local screening model~(\ref{eq:capmat}) yields poor fits.

\subsection{Dispersion relation}
\label{DispRel}

The Hamiltonian $\hat{H}_0$ can be rewritten in the form
\begin{equation}
\hat{H}_0 = \frac{1}{2} \sum \limits_{n,m =1}^{N-1} \hat{Q}_n C^{-1}_{nm} \hat{Q}_m + \frac{1}{2} \left(\frac{\hbar}{2e}\right)^2\sum \limits_{n,m=1}^{N-1} \phi_{n} L_{nm}^{-1}\phi_{m},
\label{eq:h0:ind}
\end{equation}
where we introduced the inverse inductance matrix $\widehat{L}^{-1}$ with matrix elements $L^{-1}_{nm}$, such that
\begin{equation}
 \widehat{L}^{-1} = \left( \begin{array}{ccccc c} \frac{2}{L_J} & \frac{-1}{L_J} & 0 & \hdots & &\vspace{2mm}\\
\vspace{2mm} \frac{-1}{L_J} &  \frac{2}{L_J} & \frac{-1}{L_J} &  0 & \hdots &\\ \vspace{2mm}
0 & \frac{-1}{L_J} &  \frac{2}{L_J} & \frac{-1}{L_J} &  0  & \hdots \\ \vspace{2mm}
\vdots & 0 & \ddots &  \ddots & \ddots & \ddots \\ \vspace{2mm}
& & & & & \\
\end{array} \right).
\label{eq:indmat}
\end{equation}
Here $L_J= (\hbar/2e)^2 (1/E_J)$ is the Josephson inductance.

Since $\hat{H}_0$ is quadratic, it can be straightforwardly diagonalized and represented in second quantized form,
\begin{equation}
\hat{H}_0 = \frac{1}{2} \sum \limits_{k=1}^{N-1} \hbar \omega_k \hat{a}^\dagger_k \hat{a}_k.
\label{eq:h0:secquant},
\end{equation}
Operators $\hat{a}^\dagger_k$ and $\hat{a}_k$ are bosonic; they create and annihilate excitations of the electromagnetic modes sustained by the chain. The frequencies $\omega_k$ as function of $k$ constitute the dispersion relation of these modes along the chain. They are found by solving the eigenvalue problem
\begin{equation}
\widehat{C}^{-1/2} \widehat{L}^{-1} \widehat{C}^{-1/2}  \vec{\psi}_{k} = \omega_k^2  \vec{\psi}_{k}.
\label{eq:eigp}
\end{equation}

\subsection{Weak nonlinearity and Kerr coefficients}
\label{WNL}

The eigenvectors $ \vec{\psi}_{k} $ of the matrix $\widehat{C}^{-1/2} \widehat{L}^{-1} \widehat{C}^{-1/2} $ are related to the spatial distribution of charge and phase along the chain for the corresponding eigenmode $k$. For instance, introducing the vector $\vec{\phi} = (\phi_1, \ldots, \phi_{N-1})$, the second-quantized expression for the phases $\phi_n$ along the chain is given in compact notation by
\begin{equation}
\vec{\phi} = 2e \sum \limits_k \sqrt{\frac{1}{2 \hbar \omega_k}} (\hat{a}^\dagger_k + \hat{a}_k) \widehat{C}^{-1/2} \vec{\psi}_{k}.
\label{eq:sec_quant_phase}
\end{equation}

With the help of Eq.~(\ref{eq:sec_quant_phase}), the perturbative part of the Hamiltonian $\hat{H}_1$, Eq.~(\ref{eq:h1}), can also be expressed in second-quantized form. Referring the reader to Ref.~\cite{Weissl2} for details, here we only present the result for the diagonal part of the Hamiltonian, including the fourth order correction:
\begin{align}
\hat{H}_0+H_1 ={}&{} \sum_k \hbar \omega'_k   \hat{a}_k^\dag\hat{a}_k 
- \frac\hbar{2}\sum_{k,k'}  K_{kk'}  \hat{a}_k^\dag\hat{a}_k\hat{a}_{k'}^\dag\hat{a}_{k'}+{}\nonumber\\
{}&{}+(\mbox{off-diag.}),
\label{eq:NLHam}
\end{align}
where ``(off-diag.)'' stands for fourth-order terms generated by Eq.~(\ref{eq:h1}) which contain processes including more than two photons and which are not probed in the present experiment. The coefficients $K_{kk'}$ which describe the two photon process are given by
\begin{align}
K_{kk'} &= 2(2-\delta_{kk'})\,\frac{\pi^4 \hbar E_J}{\Phi_0^4 C_J^2\omega_k\omega_{k'}}\, \eta_{kkk'k'}, \label{eq:Kkk}
\end{align}
where $\eta_{kkk'k'}$ are dimensionless mode wave function overlaps:
\begin{align}
\eta_{kkk'k'} =\sum_n \left[ \left( \sum_m \left(\sqrt{C}\widehat{C}_{n,m}^{-1/2}-\sqrt{C}\widehat{C}_{n-1,m}^{-1/2}\right)\psi_{m,k} \right)^2 \right. \nonumber\\
\times\left.\left(\sum_m \left(\sqrt{C}\widehat{C}_{n,m}^{-1/2}-\sqrt{C}\widehat{C}_{n-1,m}^{-1/2}\right)\psi_{m,k'}\right)^2\right].
\label{eq:eta}
\end{align}
For the short range model, Eq.~(\ref{eq:Kkk}) can be evaluated to the analytical formula
\begin{align}
K_{kk'} &= \left(\frac{1}{2}+\frac{\delta_{kk'}}{8}\right)\frac{\hbar^2\omega_k\omega_{k'}}{2NE_J}.
\label{eq:Kkk2}
\end{align}
As can be seen, the Kerr coefficients increase with increasing frequency. 
In comparison to a single Josephson junction the Kerr coefficients in a chain of $N$ junctions are reduced by a factor of $N$. The reason for this is that the mode wave function amplitude scales as $1/\sqrt{N}$.
The nonlinearity is consequently strongly reduced compared to a single Josephson junction. 
Physically, the effect of the weak nonlinearity is threefold. \\(i)~The linear mode frequencies $\omega_k$ are shifted to lower frequencies
\begin{equation}
\omega'_k = \omega_k - \sum_{k'} K_{kk'}/2,
\label{eq:omega-prime}
\end{equation}
where $K_{kk'}$ are the Kerr coefficients. This equation reflects the fact that the bare frequencies of the linear modes $\omega_k$ undergo a frequency downward shift due to the non-linear potential even in the absence of photons in the modes. \\(ii)~Two photons present in the same mode $k$ interact with each other, the corresponding nonlinear frequency shift determined by the self-Kerr coefficient $K_{kk}$. \\(iii)~Two photons present in different modes $k$ and $k'$ also interact with each other; the corresponding frequency shift is determined by the cross-Kerr coefficient $K_{kk'}$ for $k\neq{k}'$. The perturbative nature of these results implies that the Kerr shifts should be small compared to the unperturbed frequencies $\omega_k$. In other words, we require $\left|\sum_{k'} K_{kk'}/2\right| \ll \omega_k$.
In the following we will analyse our experimental data through this theoretical model.

\begin{figure}[h]
\includegraphics[width=8.5cm]{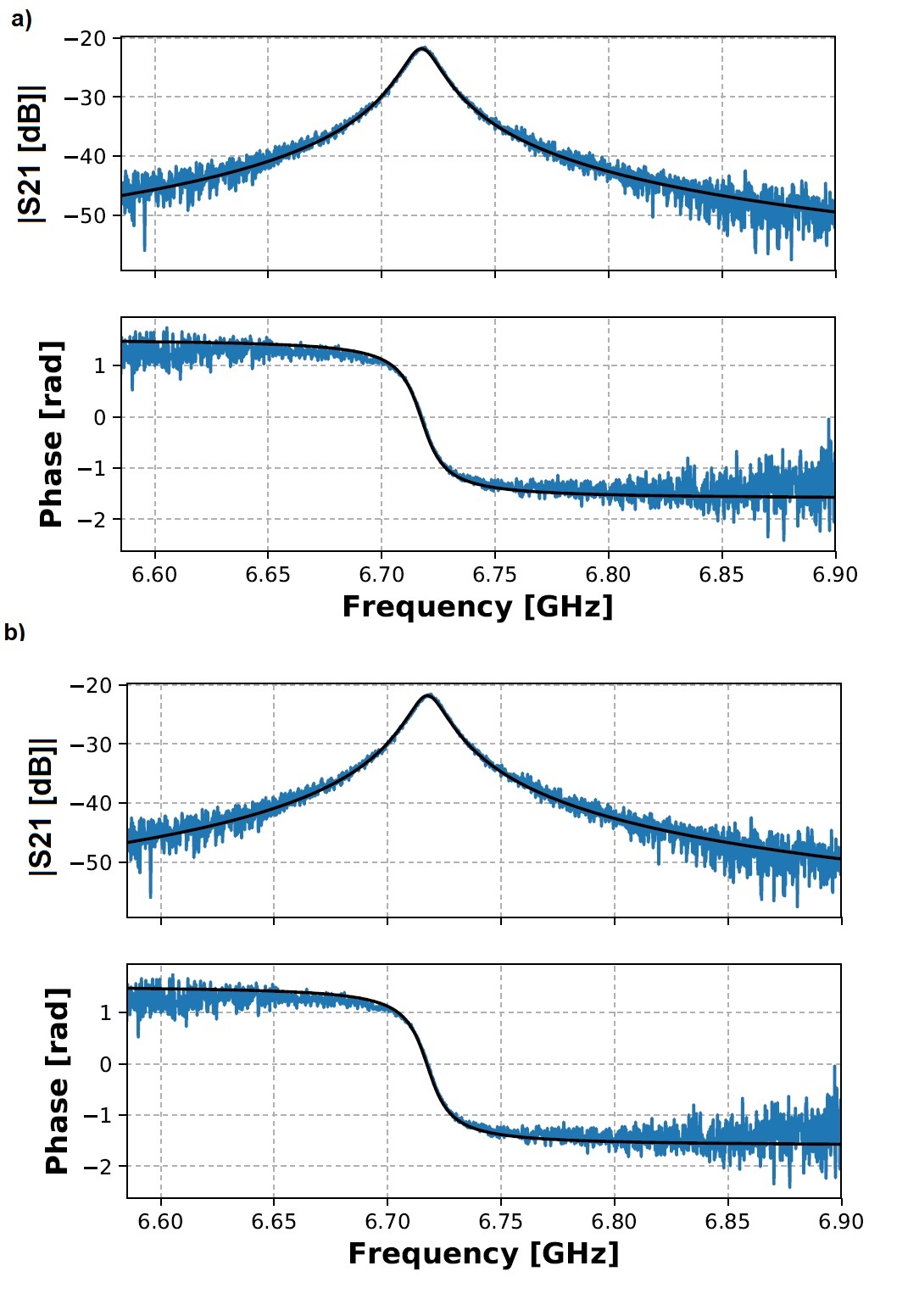}
\caption{One-tone spectroscopy of propagation modes in the chain of 500 SQUIDs: (a) 15 modes are resolved within the bandwidth of the measuring circuit 2--18 GHz; (b) Zoom of mode 3, up: amplitude, bottom: phase of the signal. Continuous line is the theoretical fit using Eq.~(\ref{eq:S21}) with $Q_i=2540$, $Q_c=535$ and $\omega_r/(2\pi)=6.717$~GHz.}
\label{one-tone}
\end{figure}

\section{Dispersion of propagating modes in a Josephson junction chain}

The transmission amplitude measured as a function of frequency, $|S_{21}(f)|$, is presented in Fig.~\ref{one-tone}~(a) within the bandwidth of 2--18 GHz of our experimental setup. We can observe directly 15 chain modes, each related to a transmission peak. Hereafter all transmission experiments were performed at zero flux. This measurement has been obtained with an input power at room temperature of $P_{in}=-60$~dBm. The attenuation of 62~dB in the input lines translates this power to an input power of $P_{sample}=-122$~dBm at the sample stage. As an example Fig.\ \ref{one-tone} (b) shows the zoom of the transmitted amplitude and phase near the frequency of mode number 3. The shape of the resonance is well fitted by the formula from Ref.~\cite{Palacious-Laloy}:
\begin{equation}
\label{eq:S21}
S_{21}(\omega) = |S_{21}(\omega)|e^{i\varphi} = \frac {1}{1+Q_c/Q_i-2iQ_{c}(\omega - \omega_r)/\omega_r}.
\end{equation}
Here the fitting parameters are the coupling quality factor $Q_c$ and the resonant frequency $\omega_r$. The internal quality factor $Q_i$ has been determined by measurements on a different set of samples shown in Appendix \ref{App:A}. 

\begin{figure}[]
\includegraphics[width=8.5cm]{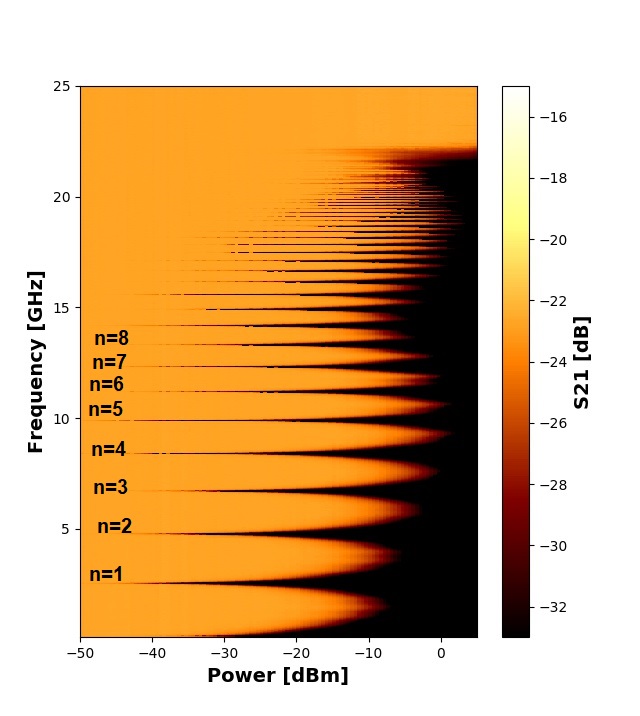}
\caption{Power-dependent two-tone spectroscopy measurement of propagation modes in the chain of 500 SQUIDs. The measurement tone for this two-tone measurement was mode $n=4$ at $8.41$GHz. The first 43 modes are clearly resolved. There is a cut-off frequency for the transmission at 22 GHz above which no signal is transmitted. }
\label{two-tone}
\end{figure}

In order to extend the measurement to higher frequencies, we apply a two-tone technique based on the cross-Kerr effect occuring in the Josephson junction chain. While the VNA measures continuously at a fixed frequency of the probe mode, the external frequency generator sweeps the pump tone frequency up to 40~GHz. When the frequency of the pumping tone hits the frequency of one of the chain modes, the probe mode exhibits the cross-Kerr frequency down-shift. This induces a dip in the measured transmission amplitude of the VNA.  Fig.~\ref{two-tone} shows a pump-power-dependent two-tone spectroscopy measurement. We can resolve clearly the 43 lowest modes of the chain. The higher modes cannot be distinguished as they merge all together just below the cut-off frequency of $\simeq 22$~GHz. 

\begin{figure}[]
\includegraphics[width=8.5cm]{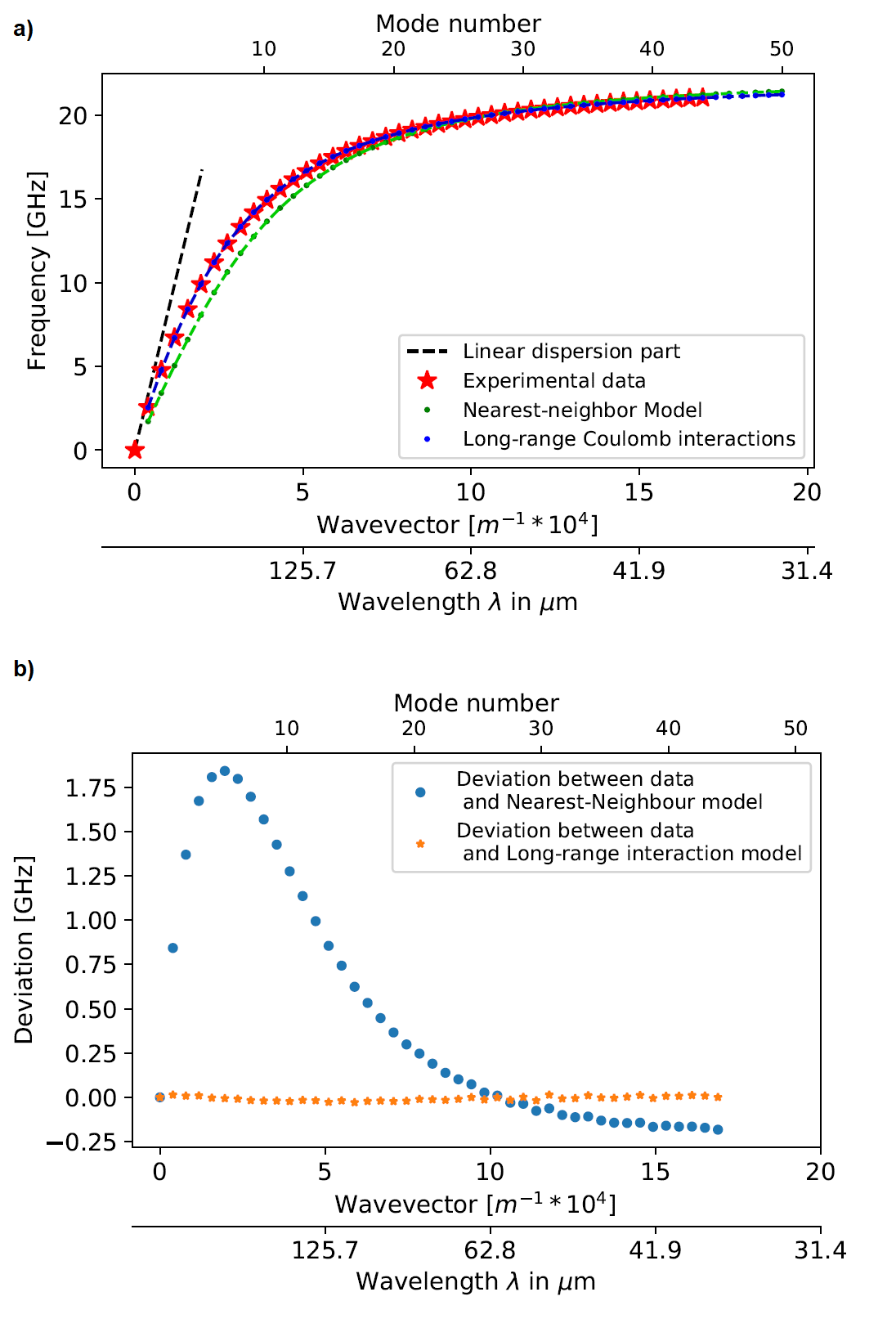}
\caption{(a)~Mode dispersion for the 500-SQUID chain. Red stars: measured data, extracted from low-power two-tone spectroscopy (see Fig.\ \ref{two-tone}); green dots: theoretical fit using the local screening model; blue dots: theoretical fit using the long-range screening model;
dashed black line: low-frequency slope.
(b)~Deviation of the mode frequencies obtained using the two models including the Kerr shifts from the measured frequencies. While the local screening model shows a maximum deviation of 25\%, the long-range screening model agrees within 0.1\% with the measured mode frequencies.}
\label{twofits}
\end{figure}

Fig.\ \ref{twofits}(a) (red stars) shows the dispersion curve deduced from the measurement shown in Fig.\ \ref{two-tone} by reading out the frequency of each mode~$m$ at its lowest possible detection power. We assume the wavevector $k_n$ for the mode number~$n$ to be given by $k=\pi{n}/L$ with $L=800\:\mu$m. In these measurements we succeed to obtain an experimental dispersion relation over a large extension on wavevector and frequency. At very low wavevector, we plotted a linear dispersion relation which can fit only the lower frequency modes. It corresponds to a refraction index of~$57$. With increasing wavevector the refraction index is even increasing more and at large wavevectors of $k=10^5\:\mbox{m}^{-1}$ the refraction index becomes as large as~250. We observe that, even in the low wavevector regime, the linear dispersion is not able to describe the experimental dispersion. To this end, we fit in Fig.~\ref{twofits}(a) the dispersion relation with the local screening model with the capacitance matrix~(\ref{eq:capmat}) and with the long-range screening model with the capacitance matrix~(\ref{eq:fullcapmat}). Both include the frequency downshift from the Kerr non-linearity~(see equation \ref{eq:omega-prime}). The long-range screening model (blue dots) fits perfectly the experimental dispersion, while for the local screening model (green dots) the calculated values of the frequencies for small wavevectors differ from the experimental ones as much as 25\% even for the best fit. This striking difference is seen especially well in Fig.~\ref{twofits}(b) where we show the respective deviation of the two models from the experimentally measured frequencies.

From the fit of the long-range screening model, we deduce the length $a_0=0.74~\mu$m, which is twice smaller than the island size of 1.6~$\mu$m. We also extract precisely the plasma frequency value $\omega_p/(2\pi) = 22.726\:\mbox{GHz}$, which gives the inductance associated with a single SQUID, $L_J = 1.56\:\mbox{nH}$. These parameters translate into a characterisitc impedance of the SQUID chain of $Z_{chain}=3.8$k$\Omega$.

\section{Self- and cross-Kerr effects in regime of weak nonlinearity}


We study experimentally the Kerr frequency shifts for seven modes of the chain with mode numbers from 2 to 8. For each selected pair of modes, the simultaneous study of self-Kerr shift of the pumped mode~$k'$ and the corresponding cross-Kerr shift of the probed mode~$k$ were performed. The measurement procedure can be explained as follows: (i)~The VNA scans the vicinity of both modes $k$ and $k'$ at low power to detect their bare frequencies; (ii)~The VNA scans mode $k'$ at higher power in order to detect its self-Kerr shift; The updated frequency for mode $k'$ as function of the input power is determined; (iii)~The external source feeds the high power to mode $k'$ at its updated frequency, while the VNA scans the mode $k$ in order to detect its cross-Kerr frequency shift. The VNA is fixed at low read-out power to ensure that the self-Kerr effect on mode k is negligeable. Then the updated frequency for mode $k$ is determined as function of the mode $k'$ power. The measurement goes on by repeating recursively steps (ii) and (iii) with gradually increasing pumping power.

\begin{figure}[]
\includegraphics[width=8.5cm]{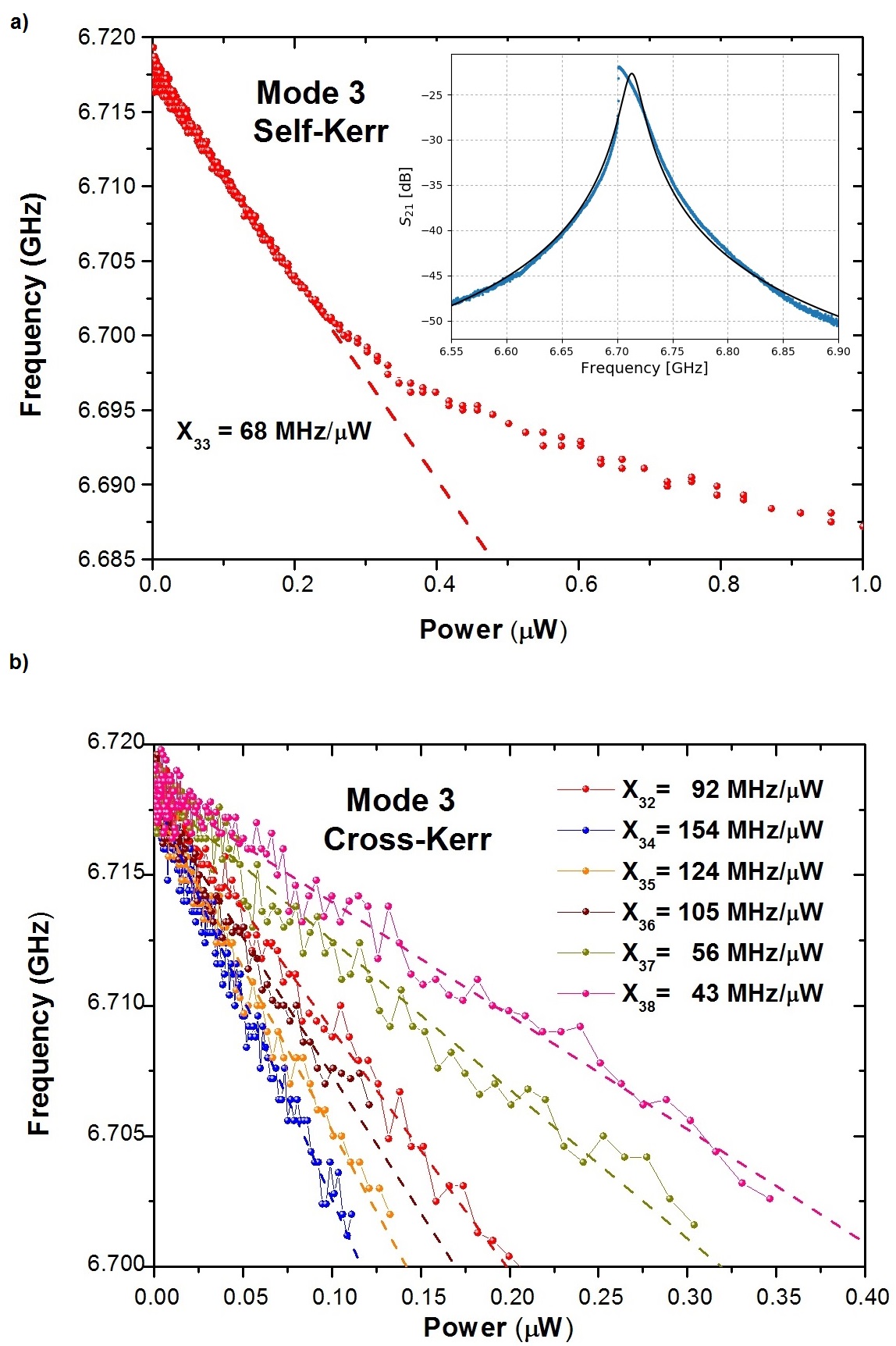}
\caption{Kerr shifts of the mode frequencies: (a) the self-Kerr frequency shift of mode $k=2$, the regimes of weak and strong nonlinearity are indicated. The inset shows the resonance at the power where bistable behavior starts to appear. From the fit of the Lorentzian we deduced $Q_{tot}=149$ and (b) Cross-Kerr frequency shifts for mode $k=3$ in the weakly nonlinear regime as a function of pumping power to a different mode (modes $k'=2,4,5,6,7$ and $8$). For each curve the corresponding values of the slope $X_{kk'}$ are extracted.}
\label{kerr}
\end{figure}

Fig.\ \ref{kerr}(a) shows typical self-Kerr shift results of these measurements. We notice a strong dependence of the frequency on the applied pump power; in particular, the frequency shift scales linearly at low pump power as expected. Above a specific power, the dependence changes. The Fig. inset shows the resonance shape measured slightly above this characteristic point: the resonance has a non-Lorentzian shape and shows a bistable state \cite{bistability}. In the following we will consider and discuss only the weak non-linearity regime for powers below this characteristic point.

Fig.\ \ref{kerr}(b) shows the cross-Kerr shifts for the mode number $k=3$. Here again, the frequency of the mode~$k$ shifts linearly with the power applied to the different modes~$k'$. Comparing the two measurements for the self- and cross-Kerr effect (shown in Fig.\ \ref{kerr}(a) and (b)), we notice a large difference in the signal-to-noise ratio for the self-Kerr and cross-Kerr traces. This difference is explained by the fact that all cross-Kerr measurements are performed at very low read-out power in order to avoid a shift in frequency due to the self-Kerr effect.

From the experiments we extract the proportinality coefficient $X_{kk'}$ which relates the input power of the mode~$k'$ to the Kerr shift on the mode~$k$. The dimensionality of these coefficients is MHz/$\mu$W, where the power corresponds to the input power on top of the cryostat.  The measured proportionality coefficients, $X_{kk'}$, are summarized in Table \ref{tab:Xjk} in Appendix~\ref{App:ExpKerrMatrix}.
In order to compare the experimentally measured coefficients $X_{kk'}$ to the theoretically calculated Kerr coefficients, strictly speaking, one has to convert the applied pump power $P_{k'}$ to the number of photons $n_{k'}$ as $n_{k'}=A_{k'}P_{k'}$, giving $X_{kk'}P_{k'}=K_{kk'}n_{k'}/2$. 
This conversion involves the mode quality factors, the attenuation of the transmission line, the external coupling strength of the mode as well as its frequency-dependent attenuation. Thus, we have to somehow deduce the experimental Kerr coefficients $K_{kk'}^\mathrm{exp}=X_{kk'}A_{k'}$ with seven unknown attenuation factors $A_2,\ldots,A_8$. We note that the matrix of Kerr coefficients must be symmetric, $K_{kk'}=K_{k'k}$. Then, the attenuation factors $A_k$ can be found up to an overall dimensional factor by minimizing the asymmetry of the resulting Kerr matrix,
\[
\min\limits_{\{A_k\}}\sum_{k,k'}\frac{X_{kk'}A_{k'}}{X_{k'k}A_{k}}.
\]
The final dimensional factor is fixed by assuming $K_{22}^\mathrm{exp}=K_{22}^\mathrm{th}$. The resulting matrix $K_{kk'}^\mathrm{exp}$ is given in Table~\ref{tab:Kexp}. The largest asymmetry in the resulting matrix still reaches $15\%$, which is of the same order as the experimental error in the off-diagonal matrix elements $X_{k\neq{k}'}$. The average asymmetry is around $4\%$.

The experimental self and cross Kerr coefficients have similar amplitude which ranges around the hundreds of kHz. They correspond to a  non-linear relative change of the refractive index $\Delta n/n\sim {10}^{-7}$ per photon.
In this sample with strong coupling to the transmission line, the Kerr coefficients are always smaller than the decay rate 
$1/T_1\simeq 60MHz$, leading to the weak coupling limit $K_{k'k}T_1\ll 1$. In the second sample with weak external capacitive  coupling, presented in Appendix \ref{App:A}, the total decay rate is weaker. The strong coupling regime is near to be achieved for the lowest mode with $K_{k'k}T_1 \sim 1$.
From the experimental parameters extracted from the dispersion relation fit, we deduce the theoretically expected Kerr matrix for our sample which is presented in Table~\ref{tab:Kjk}. 
By comparing the theoretical Kerr coefficient matrix (table \ref{tab:Kjk}) to the experimental Kerr matrix (table \ref{tab:Kexp}) we deduce an average deviation of $\Delta(K^{exp}-K^{th})\simeq 24\%$ using the following formula:
\[
\Delta(K^{exp}-K^{th})=\sum_{k,k'}(K^{exp}_{kk'}-K^{th}_{kk'})/49.
\]
 This number should be compared with our experimental precision of $14\%$.
 
By studying more carefully the experimental matrix we observe that the Kerr coefficients do not increase as 
$\omega_{k} \omega_{k'}$ but even more strikingly undergo oscillations with a typical beating of $\simeq2$ GHz as a function of mode frequency $\omega_{k}$ keeping the second frequency $\omega_{k'}$ fixed. We realised by supplementary calculations that such oscillations can be induced when the chain is coupled to resonant modes at its ends. Experimentally such modes can arise from standing modes in the injection and measurement lines. From Fig. \ref {one-tone}(a) we notice that our total transmission spectrum is superposed above 10 GHz by a beating with a frequency of around 2~GHz. This might correspond to standing waves in our measurement lines with very low quality factor (as the measurement lines are typically designed to be impedance-matched). Theoretical modeling of the influence of the non-linearity of this standing wave with strong dissipation on the measurement of our Kerr matrix goes beyond the scope of this paper. Still, we believe that the above argumentation gives a qualitative explanation why the deviation between theory and experiment is by ten percentage points larger than the experimental precision. Moreover this data-theory agreement is on par with other methods used to determine the values of self and cross-Kerr coefficients in superconducting quantum circuits~\cite{nigg_bbq}.
\begin{center}
\begin{table}
\center
\begin{tabular}{|c||c|c|c|c|c|c|c|}
\hline\hline
mode index & 2 & 3 & 4 & 5 & 6 & 7 & 8 \\
\hline\hline
2 & 8.44 & 23.87 & 21.33 & 28.81 & 23.08 & 29.60 & 24.99 \\
\hline
3 & 23.87 & 16.07 & 35.33 & 49.5 & 32.63 & 40.74 & 34.7 \\ 
\hline
4 &  21.49 & 36.45 & 20.53 & 51.57 & 38.04 & 53.16 & 38.83  \\
\hline
5 & 28.81 & 2.45 & 38.99 & 38.99 & 53 & 66.69 & 42.97 \\
\hline
6 & 23.08 & 33.9 & 39.47 & 58.41 & 22.12 & 49.5 & 26.41 \\
\hline
7 & 29.6 & 42.81 & 52.52 & 77.67 & 48.7 & 22.92 & 42.97\\
\hline
8 & 24.99 & 33.9 & 37.56 & 39.95 & 21.8 & 39.47 & 23.71  \\
\hline\hline
\end{tabular}
\caption{Experimentally determined matrix of Kerr coefficients $K_{kk'}^\mathrm{exp}$ in $10^{-2}$MHz/photon, obtained from $X_{kk'}$ in Table~\ref{tab:Xjk} by minimizing the asymmetry and fixing $K_{kk'}^\mathrm{exp}=K_{kk'}^\mathrm{th}$.}
\label{tab:Kexp}
\end{table}
\end{center}

\begin{center}
\begin{table}
\center
\begin{tabular}{|c||c|c|c|c|c|c|c|}
\hline\hline
mode index & 2 & 3 & 4 & 5 & 6 & 7 & 8 \\
\hline\hline
2 & 8.44 & 15.59 & 19.57 & 23.24 & 26.42 & 29.13 & 31.5 \\
\hline
3 & 15.59 & 16.71 & 27.69 & 32.47 & 36.76 & 40.74 & 44.09 \\
\hline
4 & 19.57 & 27.69 & 26.42 & 41.06 & 46.31 & 50.93 & 55.07 \\
\hline
5 & 23.24 & 32.47 & 41.06 & 36.61 & 54.91 & 60.32 & 65.09 \\
\hline
6 & 26.42 & 36.76 & 46.31 & 54.91 & 46.95 & 68.44 & 73.85 \\
\hline
7 & 29.13 & 40.74 & 50.93 & 60.32 & 68.44 & 56.98 & 81.33 \\
\hline
8 & 31.5 & 44.09 & 55.07 & 65.09 & 73.85 & 81.33 & 66.53 \\
\hline\hline
\end{tabular}
\caption{The matrix of Kerr coefficients $K_{kk'}^\mathrm{th}/2\pi$ in units of $10^{-2}$ MHz/photon, calculated from equations~(\ref{eq:Kkk}) and~(\ref{eq:eta}) using the long-range screening model, $L_J = 1.56\:\mbox{nH}$ and $a_0=0.74\:\mu$m.}
\label{tab:Kjk}
\end{table}
\end{center}

\section{Summary and Conclusion}
We have investigated, both experimentally and theoretically, the dispersion relation and Kerr effect of plasma modes in a one-dimensional Josephson junction chain containing 500 SQUIDs. Using the two-tone spectroscopy technique we can resolve clearly up to 43 lowest modes propagating along the chain. Remaining in the regime of weak nonlinearity, the measured dispersion curve fits perfectly with the theoretical model if we take into account two factors: 1) the bare frequencies of the modes are subjected to the Kerr non-linear renormalisation; 2) there is a long range Coulomb interaction between the island charges, resulting from the remote ground plane. To account for these long-range Coulomb interactions we introduced a remote ground model based on image charges enabling us to fit perfectly the dispersion relation without supplementary fitting parameters. From the fit of the dispersion relation we deduced the values for the plasma frequency, the inductance associated with a single SQUID and the short-range cut off length $a_0$. This enabled us to calculate the theoretical Kerr coefficent matrix. We performed measurements of the cross- and self-Kerr coefficients for the modes from 2 to 8 and compared them with our theoretical predictions. The comparison is satisfying. 
We believe that our results open new ways to design Kerr-non linearities and band-gap engineering for the realisation of Josephson parametric amplifiers and traveling-wave parametric amplifiers \cite{Planat}. More generally, our results might be used for the generation of non-classical microwave states using a Josephson junction chain as a nonlinear quantum metamaterial.

\acknowledgments
The authors thank to Ioan Pop and Juan Jose Garcia-Ripoll for fruitful discussions and to Philippe Gandit for his help in building up the cryostat used for the experiment. We also acknowledge support from the European Research council (grant 306731). The sample was fabricated in the clean rooms "Nanofab" and "PTA" (Upstream Technological Plateform). This research was supported by the ANR under contracts CLOUD (project number ANR-16-CE24-0005) and GEARED (project number ANR-14-CE26-0018).

\appendix

\section{Experimental techniques}
\label{App:ExpTech}

\begin{figure}[!]
\includegraphics[width=8.5cm]{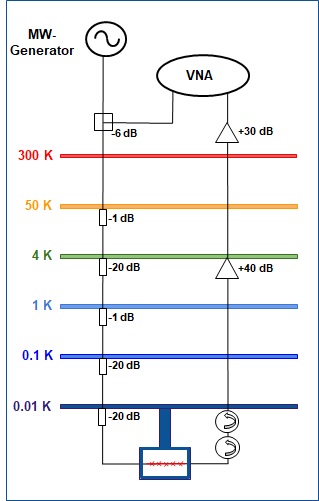}
\caption{The principal scheme of the experimental setup for one-tone- and two-tone-driven transmission measurements, performed at a temperature of $10$ mK. The transmission amplitude and phase through the SQUID-chain is measured by a Vector Network Analyzer (VNA). For mode frequencies larger than $16$~GHz we use a two-tone measurement where a second microwave tone is swept over frequency by a second microwave generator while the frequency of the VNA is kept constant. }
\label{experiment}
\end{figure}

We performed transmission measurements of the amplitude and phase with a Vector Network Analizer (VNA) in the frequency range from 2 to 40 GHz at a temperature of $10$~mK. The $50\:\Omega$ impedance coaxial transmission lines, transmitting the microwave signal to base temperature, have been step-by-step attenuated by $-62$~dB (see Fig.\ \ref{experiment}). The output line contains two amplifiers, a HEMT-amplifier at a temperature of 4K and a second amplifier at room temperature. Two circulators prevent noise emitted by the cold temperature amplifier to go back to the sample. The sample was mounted on a copper sample holder
surrounded by a black-painted copper shield. The two lowest modes of this copper cavity are $TE_{101}$ and $TE_{102}$ with frequencies of 15 and 23~GHz, respectively. No anti-level crossing is observed between the chain modes and the $TE_{101}$ mode, and the $TE_{102}$ mode is higher than the plasma frequency $\omega_p$ of the junctions. Therefore we conclude that these modes do not affect the propagating modes of the Josephson junction chain. In order to measure the response of the chain up to frequencies of $40$~GHz with the bandwidth of our experimental setup of 2--18~GHz, we use a two-tone configuration: We apply a second microwave-tone whose frequency is swept while the VNA measures the transmission at the frequency of one of the low-frequency modes of the chain.

\section{Comparison with a Josephson junction chain coupled capacitively to a transmission line}
\label{App:A}

\begin{figure}[h!]
\includegraphics[width=8.5cm]{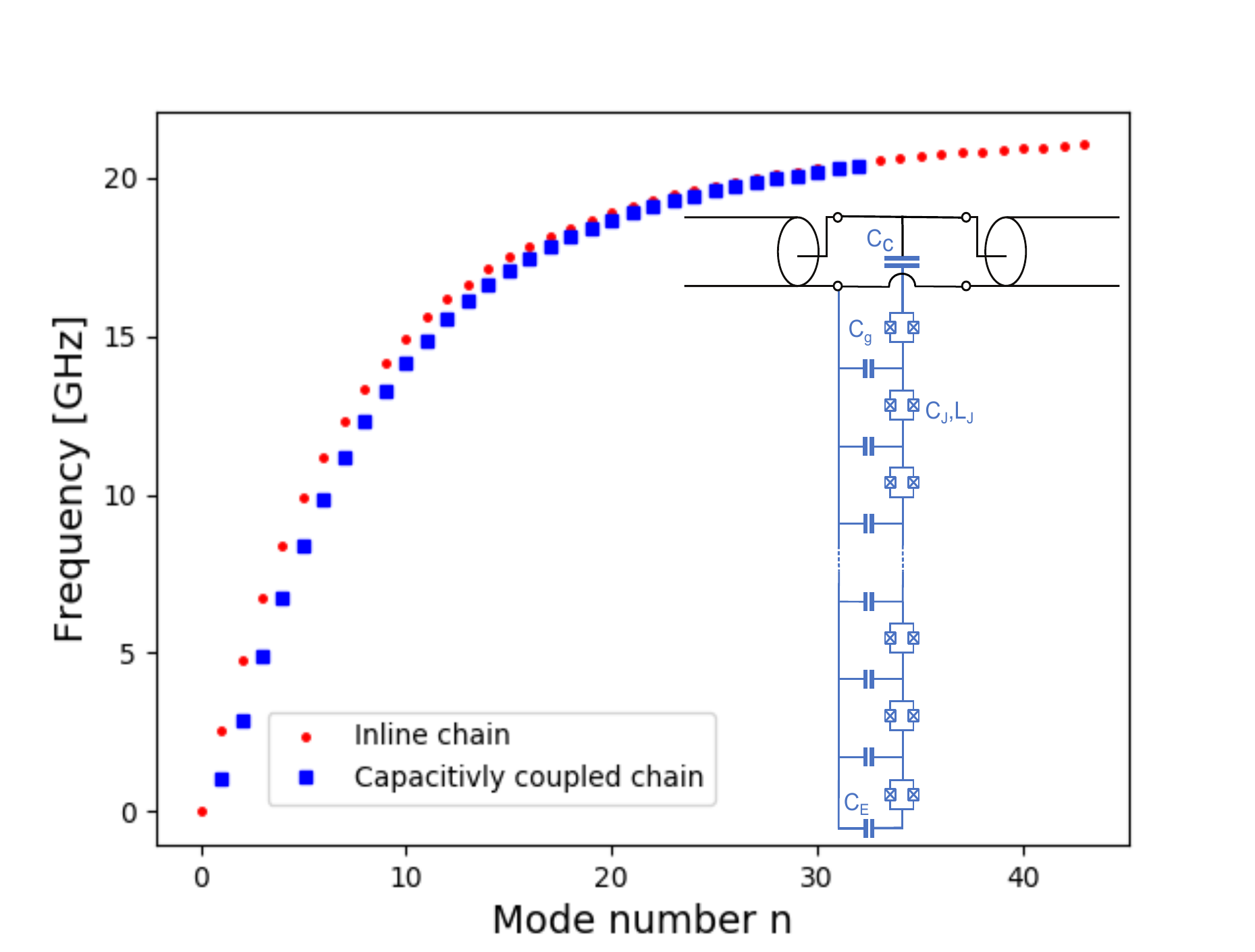}
\caption{Experimental dispersions for two identical chains of 500 SQUIDS having different coupling to the transmission line. The inset shows the equivalent scheme for the capacitively coupled chain.}
\label{appendixB}
\end{figure}

\begin{figure}[h!]
\includegraphics[width=8.5cm]{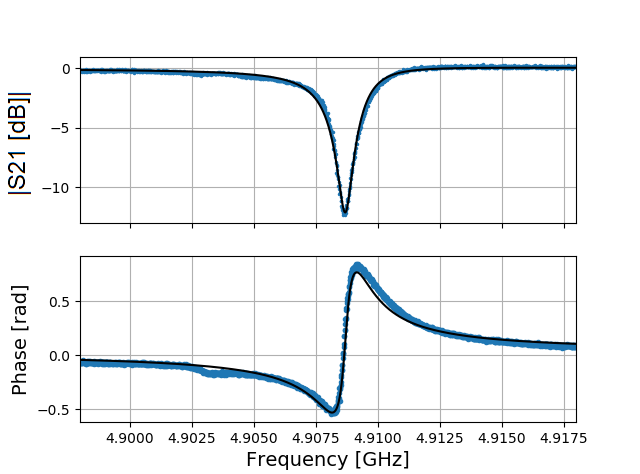}
\caption{Amplitude and phase of the resonance for the second mode. The continuous line is the fit of the resonance with equation \ref {quality_capa}. Fitting parameters are: $Q_{i}=9988$, $Q_c=3318$, $Z_0=50\:\Omega$, $X_e=10\:\Omega$ and $\omega/(2\pi)=4.909$~GHz.}
\label{appendixB2}
\end{figure}

In addition to the measurements presented in the main text, we have measured as well the dispersion relation of a 500-SQUID chain which is capacitively coupled to a transmission line as shown in the inset of Fig. \ref{appendixB}. This chain was fabricated under the same conditions with same chain parameters as the in-line chain presented in the main part of the article. In Fig.~\ref{appendixB} the dispersions of both chains are shown as a function of mode number~$n$. The value of the wavevector for the capacitively coupled chain is a more complicated expression which depends on the coupling capacitances $C_c$ and $C_E$ and has not been calculated. As the boundary conditions are different, the lowest eigenmodes with mode number $n$ have different frequencies. Both chain converge towards the same cut-off frequency, which is given by the plasma frequency $\omega_p/(2\pi)=22$~GHz. The inset of the Fig. \ref{appendixB} represents the equivalent scheme for capacitively coupled chain. Fig. \ref{appendixB2} shows the typical Lorentzian shape of a resonant mode for this chain. 

In contrast to the in-line chain, where $Q_{tot}$ is dominated by the external quality factor, the capacitively coupled chain enables us to determine the internal quality $Q_i$ of the chain and the coupling quality $Q_c$ for each chain mode. A typical resonance curve, shown in Fig. \ \ref{appendixB2}, is fitted with the formula:

\begin{equation}
\begin{split}
&S_{21}(\omega) = |S_{21}(\omega)|e^{i\varphi} \\
&S_{21}(\omega) = \frac{Z_0}{Z_0+iX_e} \frac {1+2iQ_i\frac{\omega - \omega_r}{\omega_r}}{1+\frac{Q_i}{Q_cZ_0}(Z_0+iX_e)+2iQ_i\frac{\omega - \omega_r}{\omega_r}}
\end{split}
\label{quality_capa}
\end{equation}
where $X_e$ is an asymmetry factor \cite{Dumur}. 

\begin{center}
\begin{table}[h!]
\center
\begin{tabular}{c||c|c|c|c|c|c|c|c}
\hline\hline
\# & 1 & 2 & 3 & 4 & 5 & 6 & 7 & 8 \\
\hline\hline
GHz & 2.89 & 4.91 & 6.77 & 8.42 & 9.88 & 11.18 & 12.30 & 13.28 \\
\hline
$Q_i$ & 15860 & 9980 & 2540 & 3030 & 3840 & 1150 & 1560 & 1280 \\
\hline
$Q_c$ & 1460 & 3320 & 10130 & 28800 & 14680 & 13880 & 26310 & 95000 \\
\hline\hline
\end{tabular}
\caption{Internal and coupling quality factors for modes of capacitively-coupled chain.}
\vspace{5mm}
\label{tab:quality}
\end{table}
\end{center}

\begin{figure}[h!]
\includegraphics[width=8.5cm]{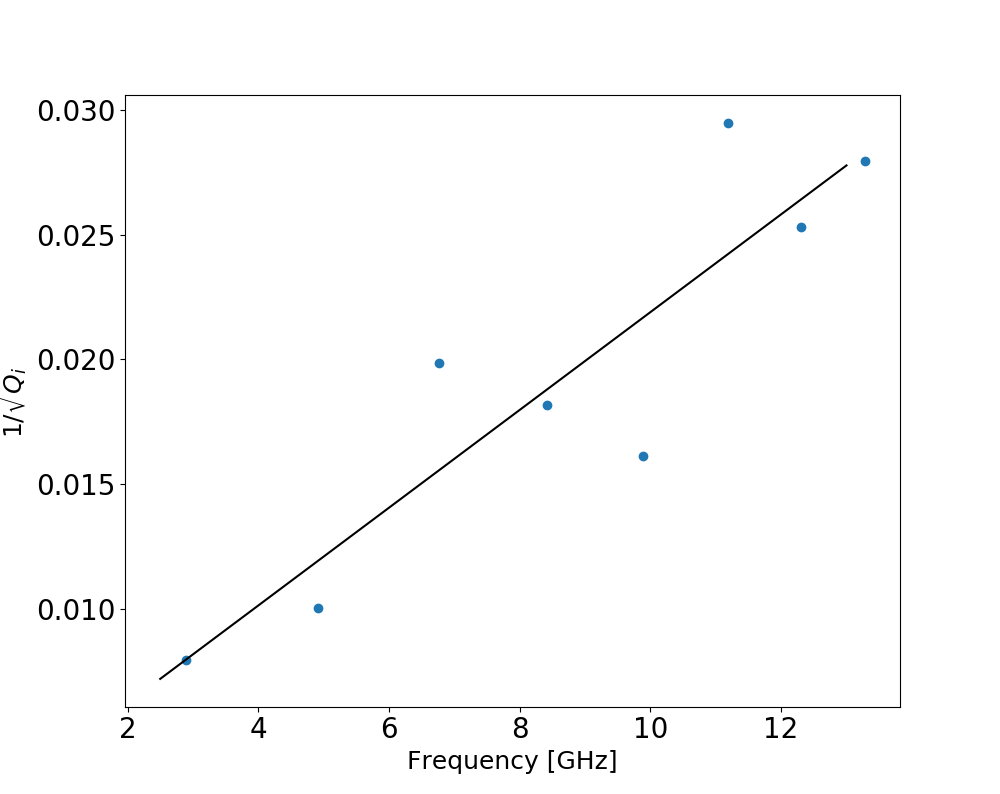}
\caption{Frequency dependance of the internal quality factor: blue points are experimental data, continous line is the theoretical fit with a $1/\omega^2$-dependance.}
\label{appendixB3}
\end{figure}
The values of the external and internal quality factor, that we obtain by fitting the resonant modes from 1 to 8, are presented in the table \ref{tab:quality}. We observe that the internal quality factor for the lowest modes can reach 15000 and decreases with increasing mode number by one order of magnitude down to 1000. The frequency dependance of the internal quality factor $1/\sqrt{Q_i}$ is shown in Fig. \ref{appendixB3}. The straight line corresponds to a $1/\omega^2$-dependence of the internal quality factor. Such a frequency dependance is expected in case of losses inside the ground plane which can be modelled by a series resistance placed between the ground capacitance $C_0$ and the ground plane.

\section{Charging energy: effect of a remote ground plane}
\label{App:ChargEnergy}

\begin{figure}[]
\includegraphics[width=8.5cm]{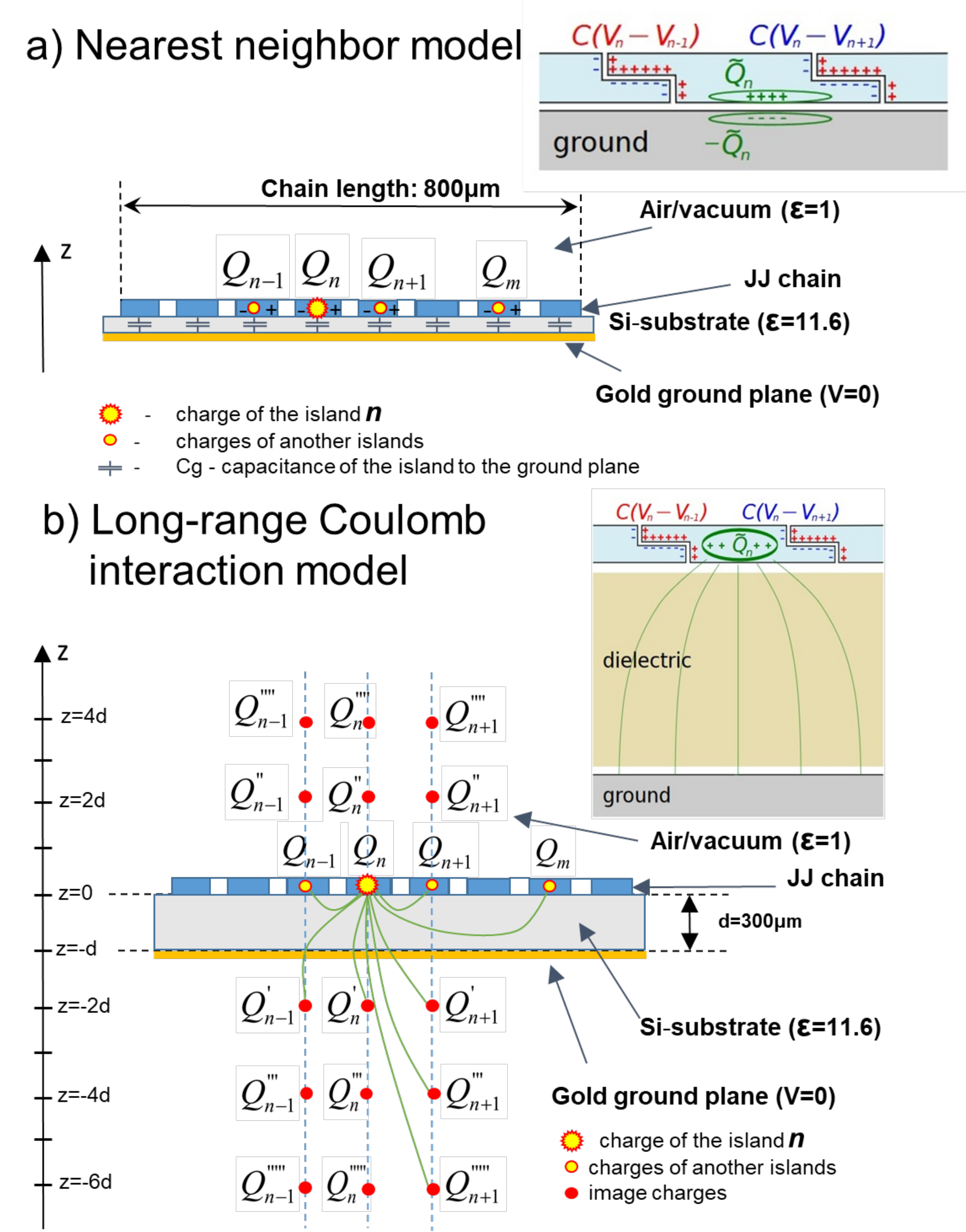}
\caption{Two different models describing the effective screening of the charge $Q_n$. (a) In the local screening model an island charge is fully screened by the ground capacitance $C_g$ and there are no long range interaction between different island charges.(b) The long-range screening model takes into account long range interactions between different island charges in the chain. For the calculation of the potential of island $n$ multiple image charges are introduced to satisfy the boundary conditions of the electrical field at the interface silicon/air and at the ground plane where $V=0$. A zoom on the single island is presented as an inset for each model.}
\label{twomodels}
\end{figure}

We first focus on the charging energy part of the Hamiltonian,
\begin{equation}
\frac{1}{2} \sum \limits_{n,m =1}^{N-1} \hat{Q}_n C^{-1}_{nm} \hat{Q}_m,
\end{equation}
which, for a given charge configuration, is completely determined by the inverse capacitance matrix $C^{-1}_{nm}$. This matrix is the response function relating the voltage $V_n$ on island $n$ to the charges $Q_{m}$ on islands $m$, $V_n = \sum_{m=1}^{N-1} C^{-1}_{nm} Q_m$, which depends on the specific dielectric environment. Following Ref.~\cite{Vogt2015}, we assume that each charge $Q_n$ residing on the corresponding island, consists of three parts (Fig.~\ref{twomodels}):
\begin{equation}
Q_n = C_J(V_n - V_{n-1}) + C_J(V_n - V_{n+1}) + \tilde{Q}_n.
\label{eq:chargeQn}
\end{equation}
The first two terms are the charges concentrated on the tunnel junctions with the neighboring islands, modeled as ideal capacitors. The charges on the opposite sides of each junction have opposite signs, so each junction is overall neutral, and does not interact with the external dielectric environment. The remaining part $\tilde{Q}_n$, unscreened by the junctions, is somehow distributed over the island, and can interact with the environment.
 Therefore the screening by the ground plane of each charge $\tilde{Q}_m$ cannot be local. Instead, one has to properly account for the long-range part of the Coulomb potential, in order to relate the voltage $V_n$ on island $n$ to the charges $\tilde{Q}_m$ on the islands $m$. For large~$d$, $\tilde{Q}_m$'s can be treated as point charges, so in the planar geometry of the present experiment, it is most natural to use the method of image charges~\cite{JacksonBook}. Our system contains two dielectric interfaces where the standard electrostatic boundary conditions on the electric field must be satisfied: the ground plane with zero potential, and the interface between the dielectric substrate and air. We consider the system, shown in Fig.~\ref{twomodels}(b). It contains a metallic plane at $z=-d$, an insulating substrate with the dielectric constant~$\varepsilon$ at $-d<z<0$, and the half-space $z>0$ is empty. 
Let us find the electrostatic potential $V(\mathbf{r})$, $\mathbf{r}=(x,y,z)$, produced by a point charge~$\tilde{Q}$, placed at the point $x'=y'=0$, $z'=0^+$ (on top of the substrate). In the two regions $-d<0<z$ and $z>0$ we seek $V(\mathbf{r})$ in two different forms:
\begin{align}
&V(x,y,z>0)=\sum_{j=0}^\infty
\frac{(4\pi\varepsilon_0)^{-1}\zeta_j\tilde{Q}}{\sqrt{x^2+y^2+(z+2jd)^2}},\\
&V(x,y,-d<z<0)=\sum_{j=-\infty}^\infty
\frac{(4\pi\varepsilon_0)^{-1}\zeta_j'\tilde{Q}}{\sqrt{x^2+y^2+(z+2jd)^2}}.
\end{align}
Indeed, each expression satisfies the Laplace equation $\nabla^2V=0$ in the corresponding region. They also must satisfy the boundary conditions at the two interfaces $z=-d$ and $z=0$. At $z=-d$ (ground plane), we have
$V(x,y,-d)=0$, which imposes $\zeta_{-j}'=-\zeta_{j-1}'$. At $z=0$, we have two conditions~\cite{JacksonBook}: first, $\partial_xV(x,y,z=0^-)=\partial_xV(x,y,z=0^+)$, which gives $\zeta_j=\zeta_j'+\zeta_{-j}'$ for $j>0$; second, $\varepsilon\,\partial_zV(x,y,z=0^-)=\partial_zV(x,y,x=0^+)$, which gives $\zeta_j=\varepsilon(\zeta_j'-\zeta_{-j}')$, again, for $j>0$. At $j=0$, we can study the solution in the limit $\mathbf{r}\to{0}$, which fixes $\zeta_0=(1+\varepsilon)/2$. This gives a closed system of equations for all $\zeta_j,\zeta_j'$; the solution for $\zeta_j$ determines the coefficients in Eq.~(\ref{eq:newV}).
 As a result, we find
\begin{align}
V_n = \sum_{m=1}^{N-1} \frac{\tilde{Q}_m}{2\pi\varepsilon_0(1+\varepsilon)}\left[
\frac{1}{\sqrt{(n-m)^2a^2+a_0^2}} \right.-\nonumber\\
-\left.\sum_{j=1}^\infty \frac{2\varepsilon(1-\varepsilon)^{j-1}/(1+\varepsilon)^j}%
{\sqrt{(n-m)^2a^2+(2jd)^2}} \right]. \label{eq:newV}
\end{align}
Here $\varepsilon \simeq 11.6$ is the dielectric constant and $a$ is the island size, defined as the total length of the chain ($800\:\mu\mbox{m}$ in our case) divided by the number of islands. The parameter~$a_0$ is a short-distance cut-off length, which must be introduced in order to avoid the divergence of the term with $m=n$ and $j=0$, representing the interaction of a point charge with itself. Clearly, at short distance, the point-like treatment of the charge~$\tilde{Q}_m$ is incorrect, and its finite spatial extent must be taken into account. Thus, $a_0$ is expected to be of the order of the island size and is treated as a fitting parameter of the model.
From the fit, we deduce the length $a_0=0.74~\mu$m, which is twice smaller than the island size of 1.6~$\mu$m.
For a finite chain, we write Eq.~(\ref{eq:newV}) in the form $V_n = \sum_{m=1}^{N-1} C^{-1}_{g,nm} \tilde{Q}_m$ and define a generalized inverse ground capacitance matrix, $C^{-1}_{g,nm}$, whose inverse, $C_{g,nm}=C_{g,mn}$, is readily calculated numerically. The matrice $C_{g,nm}$ enters into the total capacitance matrix $\widehat{C}$ of equation \ref{eq:fullcapmat} of the main text.

In an infinite chain, the dispersion relation can be found using the Fourier transform. At $d\gg{a}$, we can write
\begin{align}
&\sum_{n=-\infty}^\infty\frac{e^{-ikna}}{\sqrt{(na)^2+(2jd)^2}}
\approx\int\limits_{-\infty}^{\infty}\frac{dx}a\,
\frac{e^{-ikx}}{\sqrt{x^2+(2jd)^2}}={}\nonumber\\
&{}=\frac{2}a\,K_0(2jd|k|),
\end{align}
where $K_0(\xi)$ is the modified Bessel function. This gives
\begin{align}
&\omega^2_k=\frac{1}{LC_J}\,\frac{2(1-\cos{ka})}{2(1-\cos{k}a)
+a^2/\ell_k^2},\label{eq:omega2k=}\\
&\ell_k^2\equiv\frac{aC_J}{\pi\varepsilon_0(1+\varepsilon)}\times{}\nonumber\\
&\qquad{}\times\left[K_0(a_0|k|)-2\varepsilon\sum_{j=1}^\infty
\frac{(1-\varepsilon)^{j-1}}{(1+\varepsilon)^j}\,K_0(2jd|k|)\right].
\end{align}
Using the asymptotics  $K_0(\xi\ll{1})=\ln(2e^{-\gamma}/\xi)+O(\xi^2)$ with $\gamma=0.577\ldots$ being the Euler-Mascheroni constant, one can see that at small $k\ll{1}/d$, all logarithmic in $k$ terms cancel, so the dispersion is linear in~$k$. At $k>1/d$, the logarithmic part of the $j=0$ terms becomes important. We retrieve the dispersion relation of the one dimensional plasma modes in an homogeneous superconducting wire (Mooij-Sch\"{o}n modes \cite{Mooij}).
\begin{figure}[]
\includegraphics[width=8.5cm]{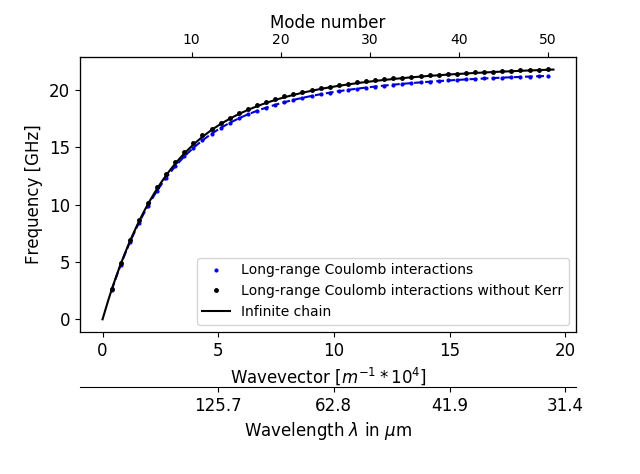}
\caption{Dispersion relation of an infinite chain (black line) compared to the propagating quantized modes (black dots) of the 500 SQUID chain without the Kerr non-linearity. Blue dots show the dispersion of the propagating modes of the 500 SQUID chain by taking into account the Kerr non-linearity.  }
\label{infinitechain}
\end{figure}

Fig.~\ref{infinitechain} shows the mode frequencies obtained from the long-range screening model without including the Kerr shifts~(\ref{eq:omega-prime}) for our 500 SQUID chain (black dots), as well as the dispersion curve for an infinite chain~(\ref{eq:omega2k=}) as a continous black line. The wavevector $k_n$ for the mode number~$n$ has been calculated assuming $k=\pi{m}/L$ where $L=800\:\mu$m. Strictly speaking, this quantization rule is not justified in the long range interaction model, for which no local boundary conditions at the chain ends can be written; indeed, the mode wave functions can deviate from a plane wave at a distance of $\sim{d}$ from the ends of the chain. Nevertheless, the dispersion of the infinite chain and the mode frequencies obtained from the long-range screening model without including the Kerr shifts agree extremely well showing that the effect of the long-range screening on the wave vector $k_n$ is negligibly small. Moreover for an infinite chain the shift to lower frequencies due to the Kerr non-linearity vanishes as can be seen from equations \ref{eq:Kkk2} and \ref{eq:omega-prime}.

\section{Experimental Kerr Matrix}
\label{App:ExpKerrMatrix}

\begin{center}
\begin{table}[!ht]
\center
\begin{tabular}{|c||c|c|c|c|c|c|c|}
\hline\hline
mode index & 2 & 3 & 4 & 5 & 6 & 7 & 8 \\
\hline\hline
2 & 32.69 & 101.43 & 93.96 & 72.54 & 74.27 & 41.45 & 31.3\\
\hline
3 & 92.28 & 68.76 & 154.36 & 124.35 & 105.35 & 56.99 & 43.48 \\
\hline
4 & 83.21 & 155.31 & 90.04 & 129.53 & 122.63 & 74.27 & 48.7  \\
\hline
5 & 111.95 & 166.4 & 208.05 & 98. & 170.98 & 93.26 & 53.91 \\
\hline
6 & 89.26 & 144.22 & 172.82 & 146.8 & 71.51 & 69.08 & 33.04 \\
\hline
7 & 114.98 & 182.25 & 229.87 & 195.16 & 157.17 & 32.09 & 53.91 \\
\hline
8 & 96.82 & 144.22 & 164.43 & 100.17 & 70.81 & 55.27 & 29.66 \\
\hline\hline
\end{tabular}
\caption{Slope coefficients $X_{kk'}$, extracted experimentally, in MHz/$\mu$W. For instance, line 3 is extracted from the slopes presented in Fig.\ \ref{kerr}(b).}
\label{tab:Xjk}
\end{table}
\end{center}

  The diagonal elements $X_{kk}$ have $\pm 6\%$ of accuracy, while the off-diagonal elements $X_{k\neq{k'}}$ show an accuracy of $\pm 14\%$, which has been deduced from the standard deviation of the linear fit of the frequency shift as a function of power.


\begin{thebibliography}{99}

\bibitem{Plourde}
B. L. T. Plourde, Haozhi Wang, Francisco Rouxinol, M. D. La Haye 
Proceedings of the SPIE 9500, Quantum Information and Computation XIII, {\bf 95000M}, (2015).

\bibitem{Eleftheriades}
G. V. Eleftheriades, A. K. Iyer, P. C. Kremer,
IEEE Transactions on Microwave Theory and Techniques {\bf 50}, 12, (2002).

\bibitem{Anlage}
S. M. Anlage
Journal of Optics {\bf 13}, 024001, (2011).

\bibitem{Pendry}
J. B. Pendry Kwak, K.-Y. Kang, Y.-H. Lee, N. Park, B. Min,
Phys. Rev. Lett. {\bf 85}, 18 (2000).

\bibitem{Jung}
P. Jung, A. V. Ustinov, S. M. Anlage,
Superconductor Science and Technology {\bf 27}, 7,(2014).

\bibitem{Alu}
A. Alu, N. Engheta,
Journal of Optics A: Pure and Applied Optics {\bf 10}, 093002, (2008).

\bibitem{Guichard}
W. Guichard, F. Hekking,  Phys Rev B,  {\bf81}, 064508 (2010).  

\bibitem{Choi}
M. Choi, S. H. Lee, Y. Kim, S. B. Kang, J. Shin, M. H. Kwak, K.-Y. Kang, Y.-H. Lee, N. Park, B. Min,
Nature {\bf 470}, 369 (2011).

\bibitem{Boulanger}
A. Dot, A. Borne, B. Boulanger, P. Seconds, C. F\'{e}lix, K. Bencheikh, J. A. Levenson, Optic Letters, {\bf 37}, 12, (2012).

\bibitem{Macklin}
C. Macklin, K. O'Brien, D. Hover, M. E. Schwartz, V. Bolkhovsky, X. Zhang, W. D. Oliver, and I. Siddiqi, {\bf 350}, 6258, pp. 307-310 Science, (2015)

\bibitem{White}
T. C. White, J. Y. Mutus, I.-C. Hoi, R. Barends, B. Campbell, Yu Chen, Z. Chen, B. Chiaro, A. Dunsworth, E. Jeffrey, J. Kelly, A. Megrant, C. Neill, P. J. J. O'Malley, P. Roushan, D. Sank, A. Vainsencher, J. Wenner, S. Chaudhuri, J. Gao, and John M. Martinis, App. Phys. Lett., {\bf106}(24), 242601 (2015)


\bibitem{Planat}
L. Planat\textit{ et al}, preprint in preparation.

\bibitem{Castellanos}
M. A. Castellanos-Beltran, K. D. Irwin, G. C. Hilton, L. R. Vale, K. W. Lehnert, Nat. Phys. {\bf 4}, 929, (2008).

\bibitem{Yurke}
B. Yurke, M. L. Roukes, R. Movshovich, A. N. Pargellis,
Appl. Phys. Lett., {\bf 69}, 3078, (1996).

\bibitem{Houck}
A. A. Houck, H. E. T\"{u}reci, J. Koch, Nat. Phys., {\bf 8}, 292, (2012).

\bibitem{Le Hur}
K. Le Hur, L. Henriet, A. Petrescu, K. Plekhanov, G. Roux, and M. Schiro, Many-Body Quantum Electrodynamics Networks: Non-Equilibrium Condensed Matter Physics with Light. arXiv.org. (2015)

\bibitem{Efetov}
K. B. Efetov, SOv. Phys. JETP, {\bf 51}, 1015 (1981).

\bibitem{Bradley}
R. M. Bradley and S. Doniach, Phys. Rev. B, {\bf 30}, 1138, (1984). 

\bibitem{Jaeger}
H. M. Jaeger, D. B. Haviland, B. G. Orr, A. M. Goldman, Phys. Rev. B, {\bf 40}, 182, (1989).

\bibitem{Giordano}
N. Giordano, Phys. Rev. Lett., {\bf 61}, 2137, (1988).

\bibitem{Chow}
E. Chow, P. Delsing, D. B. Haviland, Phys. Rev. Lett., {\bf 81}, 204, (1998).

\bibitem{Haviland}
D. B. Haviland, K. Andersson, P. Agren, J. Low Temp. Phys. {\bf 118}, 733, (2000). 

\bibitem{Duty}
K. Cedergren, R. Ackroyd, S. Kafanov, N. Vogt, A. Shnirman, and T. Duty, Phys. Rev. Lett., (2017)

\bibitem{Rastelli}
G. Rastelli, I. M. Pop, F. W. J. Hekking, Phys. Rev. B {\bf 87}, 174513 (2013). 

\bibitem{Erguel}
A. Erg\"ul, T. Wei\ss l, J. Johansson, J. Lidmar, D. B. Haviland, Sci Rep {\bf 7} 11447 (2017).

\bibitem{Pop2}
I. Pop, I. Protopopov,  F. Lecocq, Z. Peng, B. Pannetier, O. Buisson and W.  Guichard, 
Nature Physics, {\bf 6}, 589,(2010)


\bibitem{Hutter}
C. Hutter, E. A. Thol\'{e}n, K. Staningel, J. Lidmar, D. B. Haviland, Phys. Rev. B {\bf 83}, 014511 (2011).

\bibitem{Pop}
I. M. Pop, K. Geerlings, G. Catelani, R. J. Schoelkopf, L. I. Glazman, M. H. Devoret,
Nature {\bf 508}, 369, (2014).

\bibitem{Masluk}
N. A. Masluk, I. M. Pop, A. Kamal, Z. K. Minev, M. H. Devoret,
Phys. Rev. Lett {\bf 109}, 137002,(2012).

\bibitem{Altamiras}
C. Altimiras, O. Parlavecchio, P. Joyez, D.Vion, P. Roche, D. Esteve, F. Portier
Appl. Phys. Lett. {\bf 103}, 212601,(2013).

\bibitem{Bera}
S. Bera, S. Florens, H. U. Baranger, N. Roch, A. Nazir, A. W. Chin,
Phys. Rev. B, {\bf 89}, 121108, (2014).

\bibitem{Goldstein}
M. Goldstein, M. H. Devoret, M. Houzet, and L. I. Glazman, Phys. Rev. Lett. {\bf 110}, 017002 (2013).

\bibitem{Puertas}
J. P. Martinez, S. Leger, N. Gheeraert, R. Dassonneville, L. Planat, F. Foroughi, Y. Krupko, O. Buisson, C. Naud, W. Guichard, S. Florens, I. Snyman, N. Roch, arXiv 1802.00633.

\bibitem{Fistul}
S. I. Mukhin, M. V. Fistul, Supercond. Sci. Technol. 26, 084003, (2013).

\bibitem{Imamoglu}
A. Imamoglu,  H. Schmidt, G. Woods, M. Deutsch,
Phys. Rev. Lett {\bf 79}, 1467, (1997).


\bibitem{Weissl1}
T. Wei\ss l, G. Rastelli, I. Matei, I. M. Pop, O. Buisson, F. W. J. Hekking, W. Guichard,
Phys. Rev. B {\bf 91}, 014507 (2015).

\bibitem{Lecocq}
F. Lecocq, C. Naud, I. M. Pop, Z. H. Peng, I. Matei, T. Crozes, T. Fournier, W. Guichard, O. Buisson,
Nanotechnology {\bf 22}, 315302 (2011).


\bibitem{Fay}
"Couplage variable entre un qubit de charge et un qubit de phase", Aur\'{e}lien Fay, Th\`{e}se, Universit\'{e} Joseph Fourier, Grenoble (2008).

\bibitem{FazioReview}
R. Fazio and H. S. J. van der Zant,
Phys. Rep. {\bf 355}, 235 (2001).

\bibitem{Weissl2}
T. Wei\ss l, B. K\"{u}ng, E. Dumur, A. K. Feofanov, I. Matei, C. Naud, O. Buisson, F. W. J. Hekking, W. 	Guichard,
Phys. Rev. B {\bf 92}, 104508 (2015).

\bibitem{Zhang}
W. Zhang, W. Huang, M. E. Gershenson, M. T. Bell, Phys. Rev. Appl. {\bf 8}, 051001 (2017).

\bibitem{Muppalla}
P. R. Muppalla, O. Gargiulo, S. I. Mirzaei, B. Prasanna Venkatesh, M. L. Juan, L. Gr\"{u}nhaupt, I. M. Pop, G. Kirchmair
Phys. Rev. B, {\bf 97}, 024518 (2018).

\bibitem{Kuzmin}
R. Kuzmin, R. Mencia, N. Grabon, N. Mehta, Y.-H. Lin, V. E. Manucharyan, arXiv 1805.07379.

\bibitem{JacksonBook} J. D. Jackson, \textit{Classical Electrodynamics}, John Wiley \& Sons, New York (1975).

\bibitem{Bourassa}
J. Bourassa,F. Beaudoin, J. M. Gambetta, A. Blais,
Phys. Rev. A {\bf 86}, 013814, (2012).

\bibitem{Palacious-Laloy}
"Bits quantiques supraconducteurs et r\'{e}sonateurs : test de l'in\'{e}galit\'{e} de Legget-Garg et lecture en un coup", Augustin Palacious-Laloy, Th\`{e}se, Service de Physique de l'Etat Condens\'{e} CEA Saclay, (2010).


\bibitem{bistability}
R. Vijay, M. H. Devoret, I. Siddiqi, Review of Scientific Instruments, {\bf 802}, 111101, (2009).

\bibitem{Dumur}
\'E. Dumur, B Delsol, T Weissl, B. K\"{u}ng, W Guichard, C. Hoarau,
C. Naud, K. Hasselbach, K. Ratter, B. Gilles, O. Buisson,
IEEE Transactions on Applied Superconductivity, Institute of Electrical and Electronics Engineers, {\bf 26},1501304, (2016).

\bibitem{Vogt2015}
N. Vogt, R. Sch\"afer, H. Rotzinger, W. Cui, A. Fiebig, A. Shnirman, and A. V. Ustinov, Phys. Rev. B \textbf{92}, 045435 (2015).

\bibitem{Mooij}
J. E. Mooij and G. Sch\"on, Phys. Rev. Lett. \textbf{55}, 114
(1985).

\bibitem{nigg_bbq}
S. E. Nigg, et al., Black-box superconducting circuit quantization. Physical Review Letters, 108(24), 260 (2012).


\end{thebibliography}
\end{document}